\documentclass[twocolumn,aps,prl,showpacs,superscriptaddress]{revtex4}
\usepackage{color}
\usepackage{mathtools}
\usepackage{graphicx}
\usepackage{SIunits}

\begin{document}

\title{{Plasmonic Scaling of Superconducting Metamaterials}}

\author{C. Kurter}
\affiliation{Center for Nanophysics and Advanced Materials,
Department of Physics, University of Maryland, College Park MD, 20742-4111, USA}

\author{J. Abrahams}
\affiliation{Center for Nanophysics and Advanced Materials,
Department of Physics, University of Maryland, College Park MD, 20742-4111, USA}

\author{G. Shvets}
\affiliation{Department of Physics, University of Texas at Austin, Austin TX, 78712, USA}

\author{Steven M. Anlage}
\affiliation{Center for Nanophysics and Advanced Materials,
Department of Physics, University of Maryland, College Park MD, 20742-4111, USA}

\date{\today}

\begin{abstract}

Superconducting metamaterials are utilized to study the approach to the plasmonic limit simply by tuning temperature to modify the superfluid density, and thus the superfluid plasma frequency.  We examine the persistence of artificial magnetism in a metamaterial made with superconductors in the plasmonic limit, and compare to the electromagnetic behavior of normal metals as a function of frequency as the plasma frequency is approached from below. Spiral-shaped Nb thin film meta-atoms of scaled dimensions are employed to explore the plasmonic behavior in these superconducting metamaterials, and the scaling condition allows extraction of the temperature dependent superfluid density, which is found to be in good agreement with expectations.

\end{abstract}


\maketitle

$\it{Introduction.}$ There has been significant interest in the development of artificial `meta-atoms' with engineered electromagnetic properties to create qualitatively new optical media~\cite{Veselago,Smith}. These meta-atoms have dimensions much smaller than the wavelength of light at the frequency of interest, and are often arranged in periodic arrays to create an effective medium for electromagnetic waves. Early work on metamaterials was performed in the microwave part of the spectrum where it is relatively easy to create artificial structures using micron-scale fabrication techniques. It was soon realized that meta-atoms that operate well at microwave frequencies do not preserve the same properties as they are scaled down in size to operate at higher frequencies such as the infrared, visible, and ultra-violet regimes~\cite{Zhou,Ishikawa,Guney,Urzhumov,Tassin}. In particular, the ability of a split-ring resonator (SRR) and its structural derivatives to create artificial magnetism greatly diminishes and is even destroyed beyond a certain frequency range ~\cite{Zhou,Ishikawa,Guney,Dimmock,Urzhumov} mainly due to losses~\cite{Koschny}.

 For this and other reasons, a number of researchers have developed superconducting metamaterials~\cite{Ricci,Tsiatmas,Chen} which have three significant advantages over normal metal metamaterials: ultra-low losses, very compact structure (the meta-atoms can be made of deep sub-wavelength size), and the ability to be tuned and textured~\cite{AnlageJOPT}. In particular, superconductors have a strongly tunable superfluid plasma frequency~\cite{Economou}. In a metamaterial made with superconductors, one can study the approach to the plasmonic limit quite simply by tuning temperature, transport current, or magnetic field to decrease the superfluid density. In this paper we will consider the tunability of the plasmonic properties by means of temperature. We employ superconducting meta-atoms to understand the scaling dependence of their response in the plasmonic limit.

$\it{Background}$ $\it{on}$ $\it{plasmonic}$ $\it{metamaterials.}$ At low frequencies, well below the plasma frequency of a metal, it is convenient to treat the metal as a perfect electric conductor (PEC), to good approximation. In this regime, the electromagnetic response of metamaterials made of normal metals are dictated by the magnetic resonant and electric properties of PEC-equivalent structures, and essentially depend only on the geometrical shape and configuration of the meta-atoms.

At higher frequencies, approaching the plasma frequency of the metals, the PEC approximation no longer holds.  The plasma frequency of a metal depends on the carrier density ($n$) and the inertia ($m$) of the carriers as $\omega_p^2=ne^2/\epsilon_0 m$.  The complex dielectric function of the normal metal making up the meta-atoms in the vicinity of the plasma edge has the form $\epsilon_{nm}(\omega)/\epsilon_0=1-\omega_p^2/\omega(\omega+i\gamma)$, where $\omega$ is the angular frequency and $\gamma$ is the relaxation rate of the electrons, parametrizing the losses. The skin depth of the metal can be expressed in terms of the free-space wavelength $\lambda_{em}$ and the dielectric function as $\delta\sim\lambda_{em}$/$\sqrt{-Re(\epsilon_{nm}/\epsilon_0)}$. This expression shows clearly the effects of approaching the plasma edge from below \-- the screening length for electromagnetic fields systematically increases, resulting in deep penetration of fields and currents into the meta-atoms. We refer to this case as the $\textit{plasmonic limit}$.

 A resonant meta-atom in a metamaterial will store energy in the form of electric, $U_{el}$ and magnetic fields, $U_{mag}$ (both internal and external to the meta-atom), and in the form of kinetic energy of screening currents, $U_{kin}$.  The plasmonic parameter, $R_{p}$ measures the degree of plasmonic screening, or field penetration, and is defined as $R_{p}$ $\equiv$ $U_{kin}$/$U_{mag}$~\cite{Urzhumov}. In the PEC limit $R_{p}$ becomes much less than unity.  When the kinetic energy of the oscillating free electrons becomes comparable to the energy stored in the magnetic field, one can refer to the structure as being operated in the $\textit{plasmonic regime}$~\cite{Urzhumov}.  A structure enters the plasmonic regime when it is thinner than, or on the order of the skin depth $\delta$. In the plasmonic limit the persistence of artificial magnetism is very susceptible to losses~\cite{Urzhumov}, and this is the main limitation to achieving negative permeability in the visible and ultra-violet. Can we learn something about these limitations by studying superconductors in the plasmonic regime?

$\it{Superconducting}$ $\it{plasmonic}$ $\it{metamaterials.}$ The plasmonic behavior of superconductors has not been extensively explored. There was early theoretical work predicting the existence of surface plasmons on superconductors~\cite{Economou,Mooij} which have been later experimentally demonstrated by a number of groups~\cite{Buisson,Parage,Camarota,Dunmore}. Recently, extraordinary transmission through an array of subwavelength holes in a high temperature superconducting thin film has been attributed in part to plasmonic interaction between the holes.~\cite{Tman1} Moreover, superconducting plasmonic waveguides for THz radiation have been proposed.~\cite{Tman2}

Superconductors share the same evolution of their properties as normal metals in the plasmonic limit. One can define a superconducting dielectric function~\cite{Economou} as $\epsilon_{sc}(\omega,T)/\epsilon_0$=$1-(\omega^2_{ps}(T)/\omega^2)[1+i(\sigma_1(\omega,T))/(\sigma_2(\omega,T))]$ where $\omega_{ps}$ is the superfluid plasma frequency and $\sigma=\sigma_1-i\sigma_2$ is the complex conductivity. Superconductors satisfy the condition $\sigma_1/\sigma_2\ll1$  for almost all temperatures below the critical temperature, $T_c$ and frequencies such that $\hbar \omega\ll2\Delta(T)$, where $2\Delta(T)$ is the spectroscopic energy gap of the superconductor. Under these conditions the superconductor can be treated as a lossless plasmonic material to good approximation. The superfluid plasma frequency is defined as $\omega^2_{ps}(T)=(n_s(T)e^2)/(\epsilon_0 m)$, where $n_s(T)$ is the temperature dependent superfluid density.  At low temperatures where $n_s(T)$ approaches the full carrier density of the metal and $\omega_{ps}(T)$ exceeds the energy gap (typically), the superconductor approaches the PEC limit $R_{p}\ll1$.  As $T_c$ is approached, both $n_s(T)$ and $\omega_{ps}(T)$ decrease monotonically toward zero, the field penetration (measured by the magnetic penetration depth  $\lambda^2(T)=m/\mu_0 n_s(T)e^2 \sim \lambda^2_{em}/[-Re(\epsilon_{sc}/\epsilon_0)]$ and kinetic inductance increase~\cite{Meservey,AnlageAPL}, and one enters the superconducting plasmonic limit $R_{p}\gg1$. This is analogous to the approach to the $\omega_p$ from below with increasing frequency in ordinary metals.  Thus our superconducting metamaterials near $T_c$ simulate the behavior of noble metal metamaterials in the visible and ultraviolet.

However, superconductors have a frequency limitation imposed by the superconducting energy gap. For frequencies higher than $2\Delta(T)/\hbar$, the electrodynamic behavior of the superconductor approaches that of the metal in the normal state.  The magnitude of this gap approaches zero as the temperature is increased towards $T_c$ : $\Delta(T \rightarrow T_c)=0$.  For $\hbar\omega<2\Delta(T)$  and $T\ll T_c$, the plasma frequency obeys $\omega^2_{ps}(T)=\pi\frac{\sigma_n}{\epsilon_0}\frac{\Delta(T)}{\hbar}tanh(\Delta(T)/2k_BT)$~\cite{Economou,Dressel} where $\sigma_n$ is the normal state conductivity, showing that both spectral features red-shift as $T_c$ is approached from below.

$\it{Conditions}$ $\it{for}$ $\it{artificial}$ $\it{magnetism.}$ Losses in metamaterials can destroy a negative magnetic response of the structure. Artificial magnetism can be sustained only when the transit time of the guided wave along a single unit cell of the structure is less than the decay time due to the imaginary part of the dielectric function. For a metamaterial made up of normal metals, one must satisfy the condition on the relaxation rate of~\cite{Urzhumov};

\begin{equation}
\frac{\gamma}{\omega}<\frac{1}{2\pi}\frac{\lambda_{em}/a_x}{(1+R_{p})}
\end{equation} where $a_x$ is the dimension of the metamaterial unit cell in the direction of propagation of the wave.  In the plasmonic limit $R_{p}\gg1$ the right hand side of this relation decreases to the point that the inequality can no longer be satisfied for a given $\gamma/\omega$, and artificial magnetism is lost.  In the superconducting metamaterial artificial magnetism case one has a similar condition on the dissipative part of the complex conductivity,

\begin{equation}
\frac{\sigma_1}{\epsilon_0\omega}<\frac{1}{2\pi}\frac{\lambda_{em}/a_x}{(1+R_{p})}(\frac{\lambda_{em}}{2\pi\lambda(T)})^2
\end{equation}

Note that the right-hand side is the same as that for a normal metal [see Eq.~(1)], except for the additional term involving the square of the ratio of a macroscopic length (the free-space electromagnetic wavelength) to a microscopic length (the magnetic penetration depth) of the superconductor.  This large factor allows artificial magnetism to survive to a point much deeper into the plasmonic limit, as compared to normal metals. Superconducting metamaterials have the additional advantage that the ‘metamaterial parameter’ $\lambda_{em}/a_x$ can be made very much greater than unity. For instance, $\lambda_{em}/a_x \sim 10^3$ has been recently demonstared in high $T_c$ superconducting spiral meta-atoms~\cite{Ghamsari,GhamsariAPL}. An example calculation of the left and right hand sides of Eq. (2) as a function of temperature for a Nb spiral metamaterial with $a_x$= 10 mm excited at 75 MHz is shown in Fig.~1.
\begin{figure}
\centering
\includegraphics[bb=14 285 589 673,width=3.2 in]{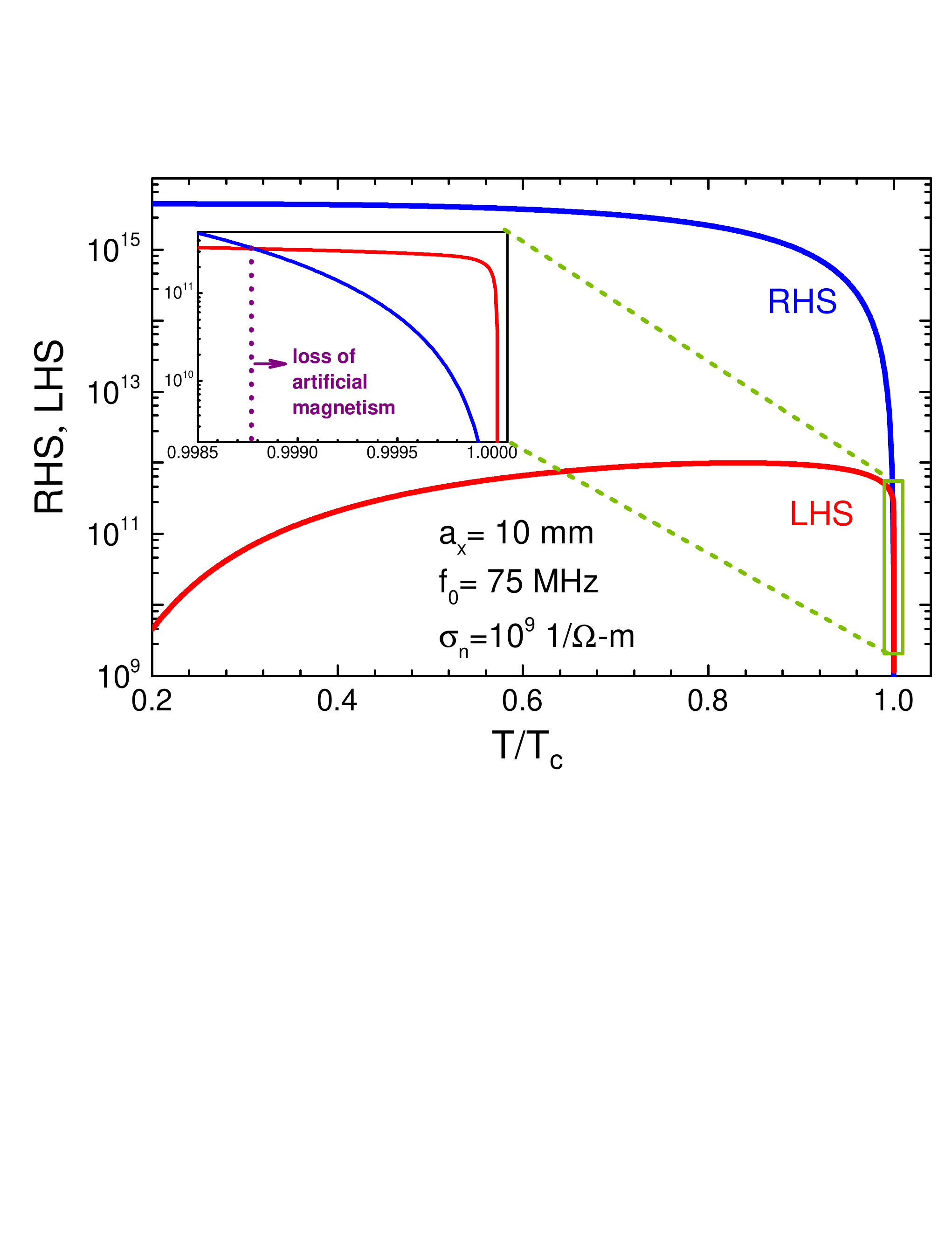}
\caption{(Color online) Plot of calculated dimension-less RHS (right-hand side) and LHS (left-hand side) of Eq.~(2) against normalized temperature. The conductivity is calculated at 75 MHz using weak-coupled Mattis-Bardeen theory~\cite{MB} in the Pippard limit.  Other parameters for the calculation are given in the figure.  The condition for artificial magnetism in a superconducting metamaterial is that RHS exceeds $\sigma_1/\epsilon_0\omega$.} \label{Fig1}
\end{figure}

$\it{Plasmonic}$ $\it{scaling.}$ Combining Ampere’s law with Faraday’s law in a dielectric medium, one can derive the harmonic wave equation $\vec{\nabla}\times\vec{\nabla}\times\vec{E}=\epsilon_r({\omega^2}/{c^2})\vec{E}$ where $\epsilon_r$ is the relative dielectric function and $c$ is the speed of light in vacuum. The scale invariance of this wave equation is governed by the quantity $\epsilon_r({\omega^2D_0^2}/{c^2})$ , where $D_0$ is a characteristic length scale.  We shall apply this scale invariance to metamaterials made up of meta-atoms of scaled dimension $D_0$ in the following way.  Consider a superconducting meta-atom having deep sub-wavelength dimensions and made up of a material with dielectric function $\epsilon_{sc}(T)$. For a given value of $\epsilon_{sc}(T)$ there is a specific value of $K\equiv\omega D_0/c$ at which resonance is achieved. Now consider another resonator with all feature dimensions scaled by a common factor with overall scaled size $D_0^{\prime}$. At frequencies much smaller than the plasma frequency $\omega_{ps}$, the resonance condition will be re-established for some new frequency $\omega^{\prime}=cK/D_0^{\prime}$ at the original value of $K$. In other words, the value of  $\epsilon_{sc}(T) K^2$ is fixed for a given resonance.

A constant value of $\epsilon_{sc}(T)$ corresponds to a fixed value of $n_s(T)/\omega^2$, for the case where $\epsilon_{sc}$ is real to good approximation. Hence by studying the resonant frequencies of scaled superconducting meta-atoms as a function of temperature, we can deduce the functional form of the superfluid density with temperature, $n_s(T)$.

$\it{Experiment.}$ We utilize superconducting spiral meta-atoms with resonant frequencies as low as 30.52 MHz at 4.3 K (see Fig.~2), putting them into the deep sub-wavelength limit ($\lambda_{em}/a_x \sim 653$). These meta-atoms were originally developed to form radio-frequency metamaterials operating below 100 MHz with outer diameter $D_0$ of 6 mm.~\cite{KurterAPL}. The spirals are made from 200 nm thick Nb thin films that were sputtered on to quartz wafers and patterned into a spiral structure by photolithography and reactive ion etching. The meta-atoms are excited by means of a coaxial transmission line that is terminated as a short-circuit loop~\cite{Ghamsari,KurterAPL} (Fig.~2(b), inset).  The RF flux is coupled to the spiral resonator, exciting it into a series of resonant modes that correspond to an integer number of half-wavelength current standing wave patterns along the length of the spiral.  The transmitted signal is picked up by a second co-axial loop on the other side of the spiral. Transmission ($|S_{21}|$) is measured on an RF network analyzer as a function of frequency at various temperatures ranging from 4.2 K up to the $T_c$ of Nb.

We prepared a series of scaled spirals, starting with a standard $D_0$= 6 mm  spiral containing 40 turns of superconducting wire 10 $\mu m$ wide, with 10 $\mu m$  gaps between neighboring wires. This basic design was scaled in all dimensions by factors of s = 0.5, 0.75, 1.25, 1.5, 2, and 2.5 to make spirals of outer diameters 3, 4.5, 7.5, 9, 12, and 15 mm, respectively [see the inset of Fig.~2(a)]. The wire widths and spacing were scaled correspondingly. The resonant frequencies of the scaled Nb thin film spirals at 4.3 K ranged from 149.58 MHz for the s = 0.5 spiral to 30.52 MHz for the s = 2.5 spiral.

\begin{figure}
\centering
\includegraphics[bb=0 2 585 730,width=3.2 in]{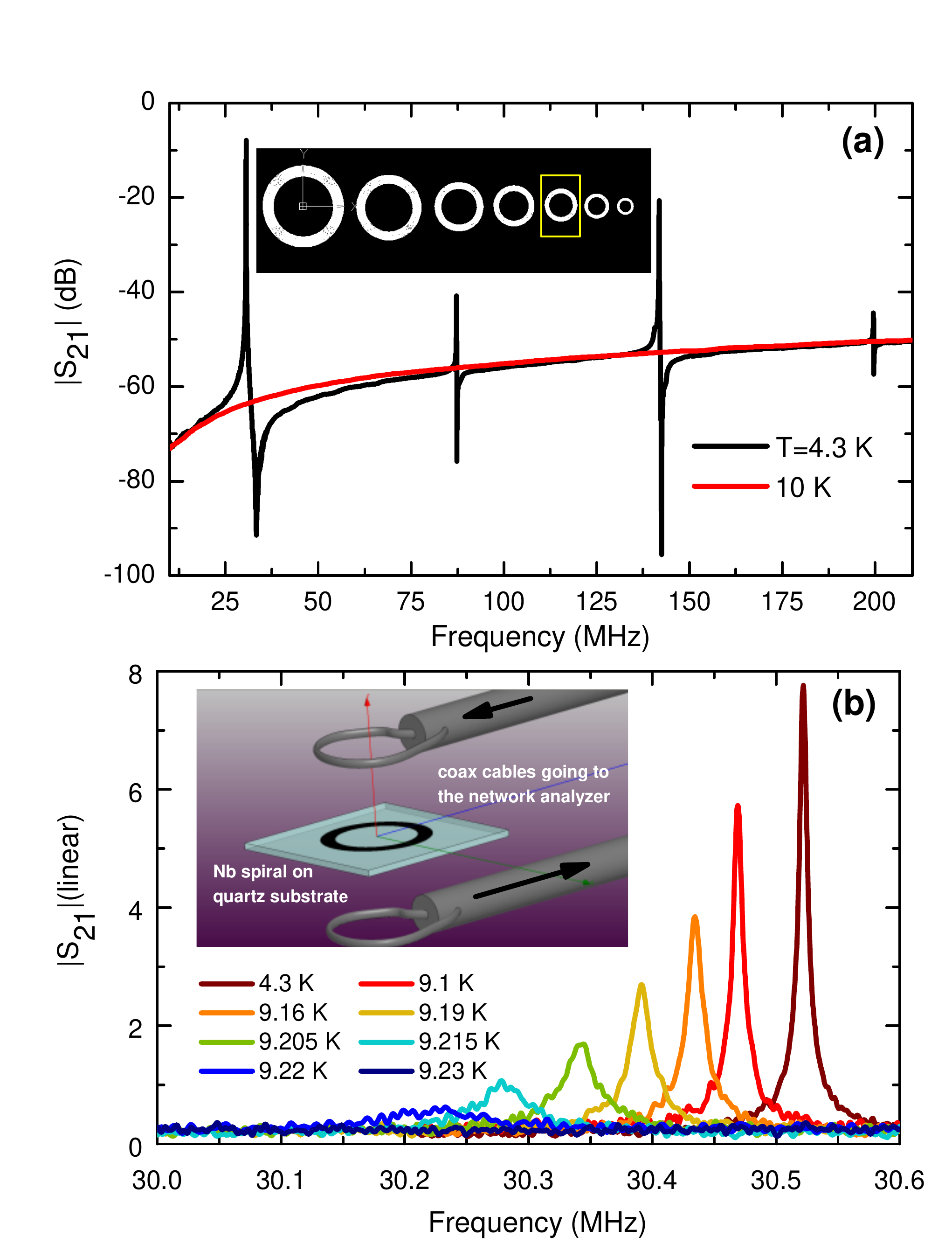}
\caption{(Color online) (a) Transmission $|S_{21}(f)|$ through one of the scaled Nb spirals with the largest scale factor (s=2.5) showing its resonance spectrum below and above the $T_c$ of Nb. The data above $T_c$ is characteristic of two loops coupled together by a mutual inductance. Below $T_c$ an additional transmission channel through the spiral appears, leading to the observed interference patterns~\cite{Ghamsari,GhamsariAPL}. The inset illustrates the scaled spiral samples; the $D_0$= 6 mm spiral is outlined with the yellow box which all others are scaled from. (b) Gradual decay of the fundamental resonant transmission peak response with increasing temperature for the same sample. The inset shows a schematic of the measurement, two coax antennas sandwiching the meta-atom.} \label{Fig2}
\end{figure}

The resonance spectrum typically shows multiple peaks [see Fig.~2(a)] although in this work we shall focus on the fundamental resonance only [Fig.~2(b)]. The modes have loaded quality factors (Q) ranging up to 5880.  The $|S_{21}|$ spectrum and mode losses for similar metamaterials are modeled in detail in refs. ~\cite{Ghamsari,GhamsariAPL}.

$\it{Results.}$ Figure 3(a) shows the temperature dependence of the normalized resonant frequency $f_0(T)/f_0(T_{min})$ versus normalized temperature T/$T_c$ for all of the scaled spirals ($0<T_{min}<<T_c$). This quantity is weakly temperature dependent well below $T_c$ because the superfluid density and the penetration depth in the superconductor saturates near its zero temperature value there~\cite{AnlageAPL}. Closer to the transition temperature, $|\epsilon_{sc}|$ decreases, the penetration depth and plasmonic parameter increase, and the spiral approaches the plasmonic limit.  As the fields penetrate more deeply into the superconductor, the magnetic and kinetic inductances increase and the resonant frequency drops~\cite{AnlageAPL}.  Note that $f_0(T)/f_0(T_{min})$ vs. T/$T_c$ varies systematically with meta-atom size $D_0$, with the smaller meta-atoms showing a stronger temperature dependence.  The inset of Fig.~3(a) shows the overall temperature dependence, where the dramatic drop of resonant frequency approaching $T_c$ is quite evident.  Finally at $T_c$ the superfluid plasma frequency $\omega_{ps}$ drops below the measurement frequency and the material reverts to the normal state, with a very large normal metal plasma frequency and correspondingly large imaginary part of the dielectric function, and the resonance is destroyed.

\begin{figure}
\centering
\includegraphics[bb=50 144 504 731,width=3.2 in]{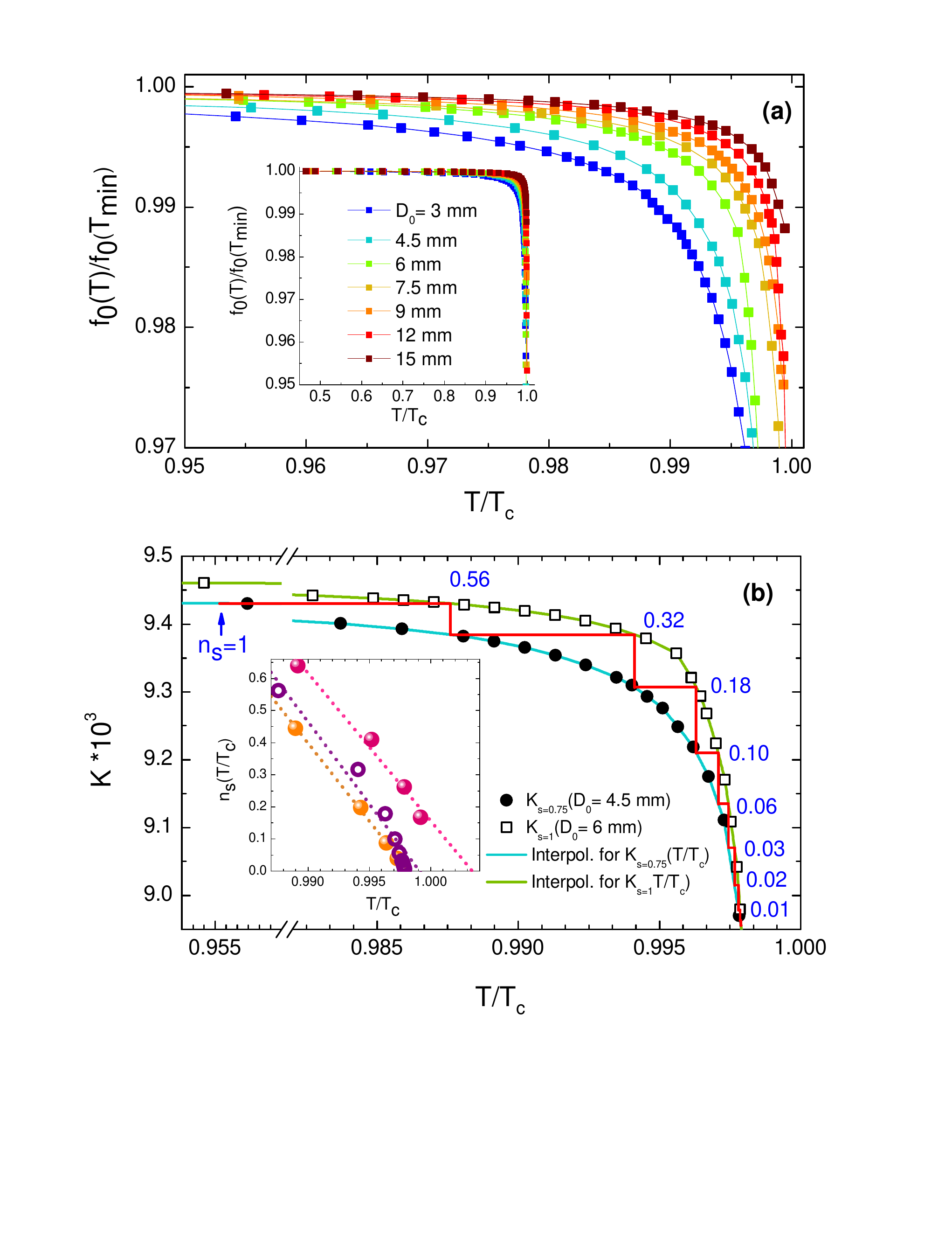}
\caption{(Color online) (a) Plot of $f_0(T)/f_0(T_{min}$) vs. $T/T_c$. Inset shows the same quantity over a larger temperature range. (b) Plot of K for two scaled Nb spirals as a function of reduced temperature $T/T_c$. The analysis to estimate $n_{s}(T/T_c)$ is shown with red stairs. Extracted values of  $n_{s}$ at specific $T/T_c$ are labelled in blue. Inset is the plot of extracted scaled superfluid density $n_{s}(T/T_c)$ versus scaled temperature  $T/T_c$ from a set of scaled Nb spiral resonators. The solid lines are the best fit to the extracted $n_{s}(T/T_c)$ for each set, demonstrating the linear trend according to $n_s(T/T_c)\sim1-T/T_c$.} \label{Fig3}
\end{figure}

As temperature varies, $\epsilon_{sc}(T)$ changes [due to changes in $n_{s}(T)$], causing the resonant frequency to vary such that the resonant condition $\epsilon_{sc}(T)K^2$ remains fixed to good approximation.  We can utilize this variation to extract the behavior of $n_s(T)$ of the superconductor making up the meta-atom.

Figure~3(b) is a representative plot of $K(T/T_c)$= $\omega(T/T_c)D_0/c$ for two scaled spirals with s = 1 and s = 0.75, having $D_0$= 6 mm and 4.5 mm respectively.  Consider first a point at temperature $T_1$ much smaller than $T_c$ on the $K_{s=0.75}(T/T_c)$ data set.  At this temperature ($T_1/T_c$) arbitrarily designates $n_s(T_1/T_c)=1$.  Now consider the point on the neighboring $K_{s=1}(T/T_c)$ curve at the same value of K. This point is at a higher temperature $T_2/T_c$.  According to the scale invariance it should have the same value of $n_s(T/T_c)/\omega^2$, such that $n_s(T_1/T_c)/\omega^2_{s=0.75}(T_1/T_c)=n_s(T_2/T_c)/\omega^2_{s=1}(T_2/T_c)$, or in other words $n_s(T_2/T_c)=n_s(T_1/T_c)[\omega_{s=1}(T_2/T_c)/\omega_{s=0.75}(T_1/T_c)]^2$, where $\omega_s(T/T_c)$ is the resonant frequency of the spiral of scale-factor s at reduced temperature T/$T_c$.  In this particular case we find $n_s(T_2/T_c)=0.56*n_s(T_1/T_c)$. Now move vertically on the K(T/$T_c$) axis to find the intersecting point on the $K_{s=0.75}(T/T_c)$ curve.  This point has the same value of $n_s(T/T_c)$.  The process can now be iterated to find the next point in reduced temperature where the superfluid density is scaled by the ratio of the resonant frequencies squared.  This process will continue until reaching a point on the  $K_{s=0.75}(T/T_c)$ curve that is below all measured points on the $K_{s=1}(T/T_c)$ curve.

The extracted superfluid density versus reduced temperature by using several pairs of scaled spiral data is shown in the inset of Fig.~3(b). Because the absolute value of the resonant frequencies are sensitive to the location of the loops stimulating the metamaterial, which vary from run to run, each extracted $n_s(T/T_c)$ data set  will have an arbitrary slope due to the variations in $K(T/T_c)$ governed by resonant frequency. The solid lines represent the best fits to the estimated $n_s(T/T_c)$ data. Although there is some scatter, we find that the data is consistent with a linear trend, as expected from mean-field Ginzburg-Landau theory near $T_c$, namely $n_s(T/T_c)\sim1-T/T_c$,~\cite{Tinkham} which is valid for Nb thin films outside of the critical regime~\cite{AnlagePRB91,AnlagePRB96}.

$\it{Discussion.}$ The fact that the recovered $n_{s}(T/T_c)$ is consistent with expectations demonstrates that our samples display a plasmonic scaling with temperature approaching the superconducting transition temperature $T_c$. The plasmonic scaling arises from the drop of the superfluid plasma frequency $\omega_{ps}$ such that it approaches our (approximately) fixed measurement frequency.

The extracted $n_{s}(T/T_c)$ deviates from a straight line near $T_{c}$. As $T_c$ is approached from below, the approximation that $\epsilon_{sc}(T)$ is purely real (in other words $\sigma_1(T)/\sigma_2(T)\ll1$) will eventually break down.  A normal fluid contribution to the electrodynamics will become significant, and destroy the condition that $n_s(T)/\omega^2$ remain fixed for a given value of $K$.  The observed deviation is consistent with the analysis of Eq.~(2) above, where the condition to retain artificial magnetism in a superconducting spiral only breaks down at temperatures quite close to $T_c$.  Note that differences in thermal contact to the samples in each run gives rise to a variation in the measured temperature, resulting in the lateral offsets of the curves in  the inset of Fig.~3(b).

$\it{Conclusions.}$ Superconducting meta-atoms are proven to be a good surrogate for normal metal meta-atoms in the plasmonic limit. They offer the convenience of measurement at approximately fixed frequency while using temperature as the plasmonic tuning parameter. Superconducting metamaterials are shown to possess very robust artificial magnetism due to their unique electrodynamic properties, as evidenced in Eq.~(2) and Fig.~1.  We have used the electrodynamic scaling behavior of deep sub-wavelength meta-atoms in the plasmonic limit to extract information about the superconducting dielectric function $\epsilon_{sc}(T)$. The temperature dependence of the estimated $n_s(T)$ is quite reasonable and consistent with Ginzburg-Landau theory.

$\it{Acknowledgements.}$ The work at Maryland is supported by NSF-GOALI ECCS-1158644 and the Center for Nanophysics and Advanced Materials. GS acknowledges support from  National Science Foundation under grant PHY-0851614.

%\bibliography{scaled SPIRALS}

\end{document}